\pdfoutput=1
\documentclass[aps,prd,amsmath,floats,floatfix, twocolumn,
superscriptaddress,nofootinbib,showpacs,longbibliography]{revtex4-1}
\usepackage[T1]{fontenc}
\usepackage[utf8]{inputenc}
\usepackage{txfonts}
\usepackage{verbatim}
\usepackage[dvipsnames, usenames, table]{xcolor}
\definecolor{linkcolor}{rgb}{0.0,0.3,0.5}
\usepackage[hypertexnames=false, unicode, colorlinks=true, linkcolor=linkcolor,
citecolor=linkcolor, filecolor=linkcolor,urlcolor=linkcolor,
pdfusetitle]{hyperref}
\usepackage{orcidlink}
\usepackage[all]{hypcap}
\usepackage{graphicx}
\usepackage{xspace}
\usepackage{amssymb}
\usepackage{microtype}
\usepackage{subfigure}

\usepackage{url}

\usepackage[english]{babel}
\usepackage{blindtext}

\usepackage[normalem]{ulem} 
\usepackage{bm} 
\usepackage{mathrsfs}
\usepackage{xspace} 
\usepackage{fontawesome}

\newcommand{\orcid}[1]{\href{https://orcid.org/#1}{\includegraphics[width=10pt]{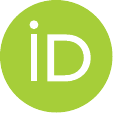}}}
\newcommand{\GWAgent}{\textcolor{linkcolor}{\texttt{GWAgent}}}
\newcommand{\pySR}{\textcolor{linkcolor}{\texttt{pySR}}}

\DeclareMathAlphabet{\mathpzc}{OT1}{pzc}{m}{it}
\usepackage[nomessages]{fp}

\newcommand{\chiS}{\chi_S}
\newcommand{\chiA}{\chi_A}

\defcitealias{Islam:2025drw}{IW25}

\begin{document}
\title{Simulation-grounded agentic AI for discovering interpretable gravitational-wave surrogates}

\author{Tousif Islam\,\orcid{0000-0002-3434-0084}}
\email{tousifislam@ucsb.edu}
\affiliation{Kavli Institute for Theoretical Physics,University of California Santa Barbara, Kohn Hall, Lagoon Rd, Santa Barbara, CA 93106}

\author{Digvijay Wadekar\,\orcid{0000-0002-2544-7533}}
\affiliation{\mbox{Center for Gravitational Physics, University of Texas at Austin, Austin, TX 78712, USA}}
\affiliation{\mbox{Department of Physics and Astronomy, Johns Hopkins University, 3400 N. Charles Street, Baltimore, Maryland, 21218, USA}}

\author{Tejaswi Venumadhav\,\orcid{0000-0002-1661-2138}}
\affiliation{\mbox{Department of Physics, University of California at Santa Barbara, Santa Barbara, CA 93106, USA}}
\affiliation{\mbox{International Centre for Theoretical Sciences, Tata Institute of Fundamental Research, Bangalore 560089, India}}

\author{Matias Zaldarriaga\,\orcid{0009-0007-8315-6703}}
\affiliation{\mbox{School of Natural Sciences, Institute for Advanced Study, 1 Einstein Drive, Princeton, NJ 08540, USA}}

\author{Ajit Kumar Mehta\,\orcid{0000-0002-7351-6724}}
\affiliation{\mbox{Chennai Mathematical Institute, Siruseri 603013, Chennai, India}}

\author{Javier Roulet\,\orcid{0000-0003-3268-4796}}
\affiliation{\mbox{Kavli Institute for Cosmological Physics, The University of Chicago, 5640 South Ellis Avenue, Chicago, Illinois 60637, USA}}
\affiliation{\mbox{School of Natural Sciences, Institute for Advanced Study, 1 Einstein Drive, Princeton, NJ 08540, USA}}

\author{Barak Zackay\,\orcid{0000-0001-5162-9501}}
\affiliation{\mbox{Department of Particle Physics \& Astrophysics, Weizmann Institute of Science, Rehovot 76100, Israel}}


\hypersetup{pdfauthor={Islam et al.}}

\date{\today}

\begin{abstract}
Fast surrogate models for expensive simulations are now essential across the sciences, but many high-performing surrogates operate as black boxes and provide limited physical insight. We present \GWAgent{}, a simulation-grounded large language model workflow for constructing interpretable analytic surrogates from numerical data. The approach is designed for settings in which candidate models can be generated, executed, and quantitatively validated against held-out simulations at each iteration. We demonstrate the method on a challenging problem in gravitational-wave (GW) physics: modeling waveforms from eccentric, spinning binary black hole mergers. Starting from a physics-informed decomposition of the waveform, \GWAgent{} searches over analytic representations, tests them against reference GW simulations, and reorganizes the fitted expressions into compact structures with identifiable physical dependence on spin, eccentricity, and frequency. The resulting surrogate achieves a median waveform mismatch, the standard GW measure of waveform error, of $6.9\times10^{-4}$ while providing an $\sim 8.4\times$ speedup in waveform evaluation, outperforming symbolic-regression and conventional machine-learning baselines on the same task. As an end-to-end astrophysical demonstration, we use the surrogate to analyze the data from the GW event GW200129 and infer an eccentricity $e_{20\mathrm{Hz}}=0.099^{+0.063}_{-0.044}$. These results show that validation-constrained language-model agents can help turn simulation data into fast, accurate, and interpretable analytic surrogates for high-precision scientific inference.
\end{abstract}

\maketitle

\noindent \textbf{\textit{\textit{Introduction: }}} Modeling complex physical systems increasingly relies on numerical simulations that are computationally expensive. Scientific inference from noisy data, however, requires a model that can be queried millions of times. Surrogate models, which approximate these expensive simulations at a fraction of the cost, have therefore become an essential tool for such analyses. Such models now underpin research across the physical sciences, including weather and climate prediction~\cite{lam2023graphcast, Bi2023AccurateMG}, molecular dynamics~\cite{Schtt2017SchNetA}, fluid mechanics~\cite{Li2020FourierNO, Raissi2019PhysicsinformedNN}, cosmology~\cite{Mancini2021COSMOPOWEREC}, and gravitational-wave (GW) astronomy~\cite{Field:2011mf,Field:2013cfa}, where they enable large-scale parameter inference and uncertainty quantification.

Machine learning (ML) has significantly advanced surrogate modeling, with deep neural networks and other flexible function approximators often achieving remarkable predictive accuracy~\cite{Raissi2019PhysicsinformedNN,Li2020FourierNO,Schtt2017SchNetA,Mancini2021COSMOPOWEREC,lam2023graphcast,Bi2023AccurateMG}. However, many of these approaches operate as black boxes, making systematic interpretation and physical understanding difficult~\cite{2018arXiv181110154R,NEURIPS2020_c9f2f917,Cranmer2023InterpretableML}. In many scientific applications, predictive accuracy alone is insufficient; researchers also seek compact, interpretable equations that expose the underlying physics while respecting known physical constraints and limiting behavior~\cite{NEURIPS2020_c9f2f917,Cranmer2023InterpretableML,Udrescu2019AIFA}. Recovering such equations becomes increasingly difficult as the complexity of the system grows. This challenge is especially acute in GW astronomy, where waveform models for merging black holes must faithfully capture highly nonlinear relativistic dynamics across broad parameter spaces while meeting the stringent accuracy required for precision scientific inference~\cite{LISAConsortiumWaveformWorkingGroup:2023arg}.

Recent advances in large language models (LLMs) have introduced a complementary paradigm for scientific ML. Beyond generating text, modern language-model agents can write code, execute computational experiments, analyze quantitative results, and iteratively refine their own implementations~\cite{2021arXiv210703374C,2022Sci...378.1092L,2024arXiv240606647Q,2023arXiv230308774O,2022arXiv220111903W,2022arXiv221003629Y,2023arXiv230204761S}. Such agentic workflows are beginning to find applications across the physical and biological sciences~\cite{VillaescusaNavarro2025TheDP,2024arXiv240515793Y,2023arXiv231006770J,2023arXiv230516291W,Wang2025AutomatedAD,Nguyen2023AstroLLaMATS,Ting2025EgentAA,Xu2025OpenSP,Gandhi2025EnhancingAA,Baker:2026xtc}. However, language models are prone to hallucination, a well-known failure mode in which they generate plausible but physically incorrect hypotheses and algorithms. This limits their use in high-precision scientific modeling~\cite{Islam:2026ona}.

Unlike many scientific discovery tasks, where assessing a proposed solution often requires substantial expert judgment, the problem of surrogate modeling however admits direct quantitative validation using held-out simulation data. Candidate models can therefore be executed, benchmarked, and iteratively refined within an automated feedback loop. This continuous grounding transforms surrogate construction into an optimization problem that is naturally suited to agentic workflows. More broadly, reliable scientific use demands workflows in which candidate solutions are grounded by rigorous quantitative validation rather than linguistic plausibility alone; indeed, LLM capabilities have advanced most rapidly in domains such as mathematics and programming, where a candidate solution can be automatically verified, the setting that underpins reinforcement learning with verifiable rewards (RLVR)~\cite{DeepSeekAI2025R1, Lambert2024Tulu3}.

\begin{figure*}
	\centering
	\includegraphics[width=\linewidth]{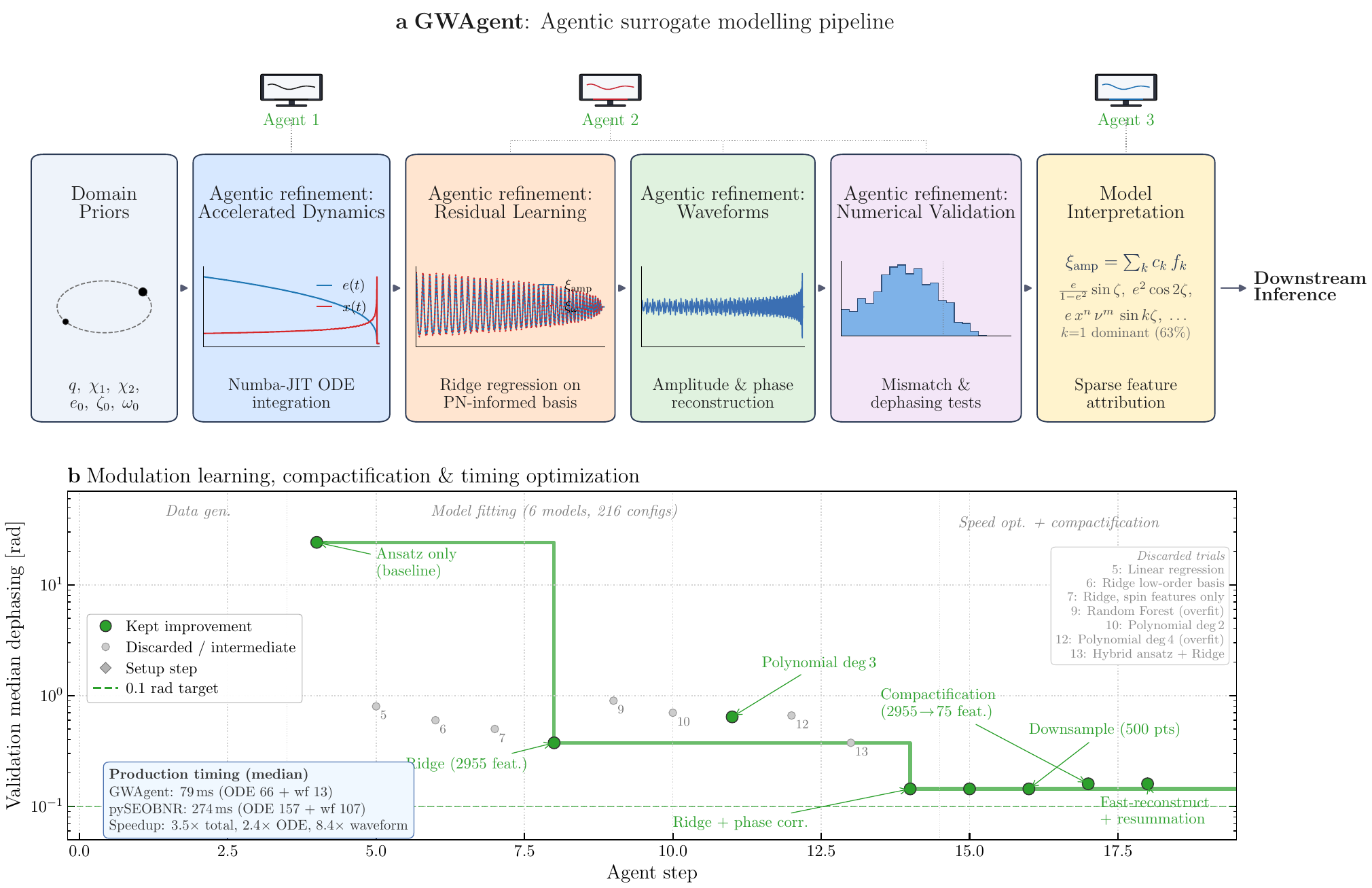}
    \caption{\textbf{Construction and optimization of analytic surrogate models for eccentric binary black hole mergers.} 
    (A) Overview of the GWAgent framework. The upper schematic summarizes the iterative workflow linking physics priors, refinement, simulation validation, surrogate construction, and astrophysical inference. Agent 1 maps binary parameters to evolving orbital variables. Agent 2 fits an analytic surrogate by learning corrections to a physics-informed baseline (see Eqs.~\ref{eq:quasicircular_ratio},~\ref{eq:residual_xi}). Agent 3 then extracts compact, interpretable structure to develop an analytic surrogate that can accurately reconstruct waveforms and be benchmarked against numerical simulations.    
    (B) We show a summary of the optimization steps performed by Agents 2 and 3, including major discarded trials that illustrate the breadth of the search space explored. Similar to evolutionary coding-agent strategies such as AlphaEvolve~\cite{Novikov2025AlphaEvolve}, the workflow accepts changes that improve accuracy or speed while retaining successful and failed attempts as context for later iterations. Progressive refinement reduces waveform phase error relative to the initial baseline while simultaneously improving efficiency, achieving phase errors close to 0.1 rad with a $\sim$8.4$\times$ speedup in waveform evaluation.}
    \label{fig:workflow}
\end{figure*}

Motivated by these considerations, we present \GWAgent{}~\footnote{\href{https://github.com/tousifislam/GWAgent}{https://github.com/tousifislam/GWAgent}}, a simulation-grounded agentic framework for constructing interpretable analytic surrogate models from expensive numerical simulations. Instead of optimizing a fixed ML architecture, it iteratively proposes analytic models, validates them against held-out simulations, and then refines the models to minimize the mismatch. We demonstrate this approach on one of the most demanding surrogate-modeling problems in computational astrophysics: GW signals from eccentric binary black hole (BBH) mergers. Unlike the nearly circular binaries observed in most GW events, eccentric binaries exhibit rapidly varying amplitude and phase modulations that substantially increase waveform complexity, making both physical modeling and surrogate construction considerably more challenging. They are also astrophysically important because different binary formation channels predict distinct eccentricity distributions, making eccentricity a key probe of binary formation and evolution~\cite{Rodriguez:2017pec,Rodriguez:2018pss,Samsing:2017xmd,Zevin:2021rtf,Samsing:2020tda}. Recent analyses further suggest that measurable eccentricity may already be present in several observed GW events~\cite{Romero-Shaw:2020thy,Gayathri:2020coq,Gamba:2021gap,Ramos-Buades:2023yhy,Gupte:2024jfe,Morras:2025xfu,romeroshaw2025gw200208222617eccentricblackholebinary,Tiwari:2025fua,Kacanja:2025kpr,Jan:2025fps,Phukon:2025cky,McMillin:2025hof}. 

Starting from a physics-informed decomposition of the waveform, \GWAgent{} jointly explores different analytic representations, basis functions, and implementation strategies to identify compact surrogate models that remain both accurate and physically interpretable. The resulting surrogate achieves a median waveform mismatch of $6.9\times10^{-4}$ together with an $\sim8.4\times$ speedup in waveform evaluation, outperforming conventional symbolic-regression and classical ML approaches on the same task. As an end-to-end scientific demonstration, we apply the surrogate to Bayesian parameter estimation of the GW event GW200129 and infer an eccentricity of $e_{20\,\mathrm{Hz}}=0.099^{+0.063}_{-0.044}$. More broadly, our results demonstrate that simulation-grounded language-model agents can discover accurate, efficient, and physically interpretable surrogate models, illustrating a general framework for trustworthy scientific AI across simulation-driven disciplines.\\

\begin{table}[t]
\centering
\setlength{\tabcolsep}{2pt}   
\caption{Comparison between \GWAgent{} and existing approaches for scientific surrogate construction.}
\label{tab:comparison}
\begin{tabular}{lccc}
\hline
\textbf{Method} & \textbf{Search space} & \textbf{Physical}& \textbf{Interpretable}\\& & \textbf{constraints} &  \\
\hline
Deep learning &
\begin{tabular}[c]{@{}c@{}}
Network\\
weights
\end{tabular}
& No & No \\
Symbolic regression &
\begin{tabular}[c]{@{}c@{}}
Symbolic\\
expressions
\end{tabular}
& Limited & Yes \\
AutoML &
\begin{tabular}[c]{@{}c@{}}
Architectures\\
and hyperparameters
\end{tabular}
& No & Limited \\
\GWAgent{} &
\begin{tabular}[c]{@{}c@{}}
Physics\\
representations
\end{tabular}
& Yes & Yes \\
\hline
\end{tabular}
\end{table}

\noindent \textbf{\textit{GW surrogate modeling:}}
GW signals from BBH mergers are naturally represented as complex time series and decomposed into spin-weighted spherical harmonic modes,
\begin{align}
h(t,\theta,\phi;\boldsymbol{\lambda})
=
\sum_{\ell=2}^{\infty}
\sum_{m=-\ell}^{\ell}
h_{\ell m}(t;\boldsymbol{\lambda})
\,{}_{-2}Y_{\ell m}(\theta,\phi),
\label{hmodes}
\end{align}
where $\boldsymbol{\lambda}$ denotes the intrinsic binary parameters and $(\theta,\phi)$ specify the observer direction. Throughout this work we consider aligned-spin eccentric BBHs described by
\[\boldsymbol{\lambda}=\{q,\chi_1,\chi_2,e_0,\zeta_0,\omega_0\},\]
where $q$ is the binary mass ratio, $\chi_{1,2}$ are the dimensionless component black hole spins, and $e_0$ and $\zeta_0$ denote the eccentricity and orbital anomaly, respectively, specified at a reference orbital frequency $\omega_0$. 
Each waveform mode is conveniently written in amplitude--phase form, $h_{\ell m}(t)=A_{\ell m}(t)e^{i\phi_{\ell m}(t)}$, with instantaneous angular frequency $\omega_{\ell m}(t)=\frac{d\phi_{\ell m}(t)}{dt}$.

Unlike typical scientific surrogate models that predict a small number of scalar quantities, GW waveform modeling requires learning high-dimensional, temporally correlated functions whose amplitudes and frequencies evolve coherently over thousands of time steps. Because the waveform phase is obtained by integrating the instantaneous frequency, small local frequency prediction errors accumulate into large phase errors, making the predicted waveforms unsuitable for precision GW inference. Eccentric orbital motion further increases the complexity of the learning task by introducing non-stationary oscillatory structure that evolves continuously throughout the binary evolution while simultaneously changing the signal duration and time sampling. 

Existing waveform-generation methods for eccentric spinning BBH mergers include a limited number of high-fidelity numerical-relativity (NR) simulations~\cite{Scheel:2025jct,Healy:2022wdn}, approximate relativistic simulations based on the effective-one-body (EOB) framework~\cite{Damour:2004bz,Konigsdorffer:2006zt,Memmesheimer:2004cv,Cho:2021oai,Cao:2017ndf,Chiaramello:2020ehz,Albanesi:2023bgi,Albanesi:2022xge,Riemenschneider:2021ppj,Ramos-Buades:2021adz,Liu:2023ldr,Wang:2023ueg,Carullo:2023kvj,Nagar:2021gss,Morras:2025nlp,Nagar:2024oyk,Ramos-Buades:2026kbq}, and phenomenological waveform models informed by post-Newtonian (PN) theory~\cite{Hinderer:2017jcs,Huerta:2016rwp,Huerta:2017kez,Tanay:2016zog,Gamboa:2024hli,Morras:2025nbp,Hinder:2017sxy,Joshi:2022ocr,Planas:2025feq,Chattaraj:2022tay,Paul:2024ujx,Manna:2024ycx,Maurya:2025shc}. Because of the additional complexities introduced by eccentric orbital motion, surrogate techniques developed for quasi-circular waveforms, including Gaussian process regression and reduced-order interpolation, become considerably more difficult to apply to eccentric BBHs~\cite{Field:2011mf,Varma:2018aht,Varma:2019csw,Islam:2021mha,Nee:2025nmh,Maurya:2025shc}.

Because relatively few spinning eccentric NR simulations currently exist, we construct an interpretable analytic surrogate using waveforms generated by state-of-the-art EOB simulations. We model the dominant GW mode $h_{22}(t)$ over a parameter space spanning the binary mass ratio, aligned component spins, initial eccentricity, anomaly, and initial orbital frequency. Reference waveforms are generated on disjoint training and validation sets that efficiently sample this parameter space, enabling an unbiased assessment of surrogate performance (Appendix~\ref{sec:app_param}).\\

\begin{figure*}
	\centering
	\includegraphics[width=\linewidth]{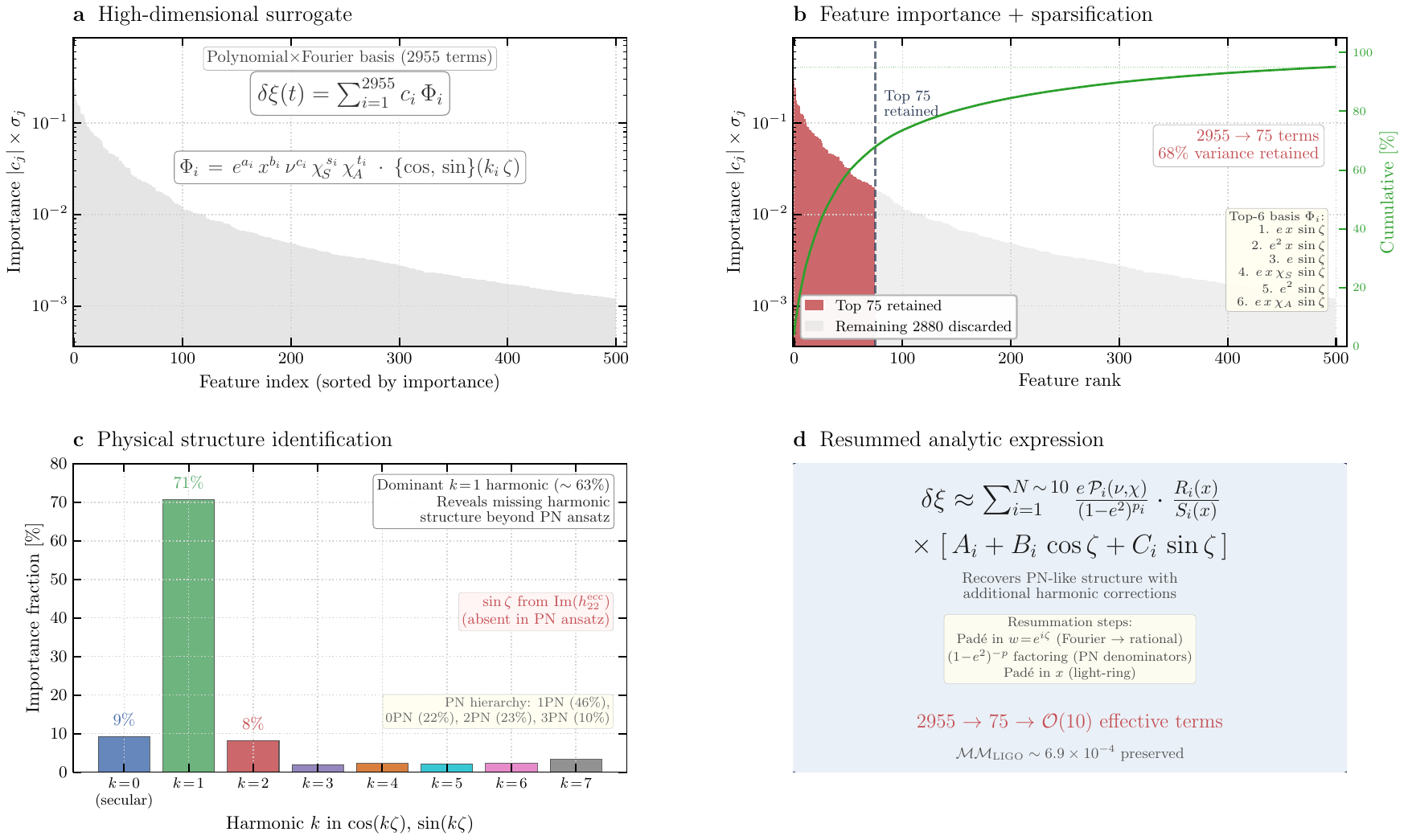}
    \caption{\textbf{Simplification and interpretation of analytic surrogate models.}
    Agent-2 from Fig.~\ref{fig:workflow} learns a high-dimensional surrogate of the residual waveform (see Eq.~\ref{eq:residual_xi}) with $\sim 3000$ basis terms. We show here the workflow of Agent-3, which ranks coefficients to identify dominant contributions, uses orbital-cycle decomposition to reveal harmonic structure and orbital-frequency scaling, and then uses the resummation technique to rewrite the result into a compact expression with $\mathcal{O}(10)$ effective terms while preserving almost the same accuracy ($\mathcal{MM}_{\mathrm{LIGO}} \sim 6.9\times10^{-4}$).}
    \label{fig:interpretation}
\end{figure*}

\noindent \textbf{\textit{Simulation-grounded agent workflow: }}Figure~\ref{fig:workflow} summarizes the \GWAgent{} workflow. The workflow consists of three stages. First, the agent constructs an efficient representation of the binary's orbital evolution, including the evolving eccentricity $e(t)$, anomaly variable $\zeta(t)$, and dimensionless orbital frequency. This optimization accelerates the underlying dynamics and waveform generation code while preserving agreement with the reference simulator (Appendix Fig.~\ref{fig:supp_dynamics_progress}). Second, the agent constructs analytic surrogate models for the eccentric waveform corrections. Finally, the fitted expressions are analyzed to identify recurring physical structure, orbital-frequency scaling, and compact analytic representations.

As our aim is to generate analytic models, we cannot use the gradient-based optimization techniques used in conventional ML workflows. \GWAgent{} improves its models through iterative interaction with an external simulator. Candidate implementations are executed, evaluated on held-out simulations using waveform mismatch and phase-error metrics, and retained only when they satisfy predefined accuracy and consistency criteria (Appendix~\ref{sec:app_metrics}). The workflow therefore follows a propose--evaluate--refine loop in which quantitative scientific validation, rather than linguistic plausibility, guides subsequent iterations. Figure~\ref{fig:workflow}b illustrates representative optimization trajectories explored by the agent, highlighting the substantially broader search space accessible through automated exploration than is practical through manual iteration.

Instead of relying on a single natural-language prompt, the workflow is initialized from a structured research protocol that specifies the scientific objective, simulation tools, target variables, validation criteria, performance targets, and expected outputs. The protocol is implemented within the open-source \GWAgent{} framework, which orchestrates iterative code generation, execution, validation, and refinement using frontier coding agents in a sandboxed environment. Unless otherwise stated, all results presented in this work were obtained using \textit{Claude Opus 4.6} through the \textit{Claude Code interface}. This separation between scientific objectives and implementation improves reproducibility by allowing the same workflow to be executed with different language models, regression algorithms, or scientific objectives under a common validation procedure (Appendix~\ref{sec:app_agentic}, Table~\ref{tab:agent_human_roles}). The complete \GWAgent{} framework, workflow protocol, and source code are publicly available at \url{https://github.com/tousifislam/GWAgent}.\\

\noindent \textbf{\textit{Physics-informed residual learning:}} Initial attempts to model the complete eccentric waveform directly did not achieve the accuracy required for GW inference. Rather than learning the full waveform from scratch, we instead ask the agent to model only the physical corrections that remain after subtracting progressively simpler, well-understood descriptions of the signal. This residual-learning strategy substantially reduces the complexity of the learning problem while preserving a direct physical interpretation of the learned quantities.

First, we compare each eccentric waveform with its quasi-circular counterpart, whose dynamics are well understood and inexpensive to evaluate~\cite{Islam:2024rhm,Islam:2024bza},
\begin{equation}
\xi_X(t)
=
\frac{X_{\rm eccentric}(t)}
{X_{\rm quasi\text{-}circular}(t)}
-1,
\label{eq:quasicircular_ratio}
\end{equation}
where $X$ denotes either the waveform amplitude or instantaneous frequency. We then subtract a physics-informed baseline $f_{\rm ansatz}(t)$ that captures the dominant eccentric modulation~\cite{Gamboa:2024hli},
\begin{equation}
\delta\xi_X(t)=\xi_X(t)-f_{\rm ansatz}(t),
\label{eq:residual_xi}
\end{equation}
leaving the surrogate to model only the remaining corrections.
This decomposition is central to the agentic workflow. The quasi-circular waveform and the leading eccentric modulation are already described by established physical models, leaving only higher-order corrections to be learned from simulation data. This residual is substantially smoother than the original waveform, simplifying optimization while preserving interpretability. The learned corrections can be directly associated with orbital cycles, frequency scaling, spin dependence, and phase drift.

The agent models $\delta\xi_X(t)$ using the evolving orbital variables together with the intrinsic binary parameters, but is not instructed which regression algorithm, basis functions, or analytic representation to use. Instead, it iteratively proposes candidate model classes, fits them to simulation data, and refines subsequent proposals using the remaining validation errors. Similar in spirit to evolutionary coding agents such as AlphaEvolve~\cite{Novikov2025AlphaEvolve} and approaches such as autoresearch\footnote{\url{https://github.com/karpathy/autoresearch}}, successful implementations are retained as context for future iterations while unsuccessful attempts help guide subsequent exploration. The final surrogate adopts a polynomial--Fourier basis in eccentricity, orbital frequency, mass ratio, spin combinations, and orbital anomaly, yielding a ridge-regression model with seven orbital harmonics and 2955 basis functions (Fig.~\ref{fig:interpretation}a).

Because the surrogate is analytic, known physical constraints can be imposed exactly. In particular, continuity with the quasi-circular limit is enforced through
\begin{align}
\lim_{e\rightarrow0}\delta\xi_{\rm amp}(t)&=0,\qquad
\lim_{e\rightarrow0}\delta\xi_{\omega}(t)=0,\qquad
\lim_{e\rightarrow0}f_{\rm ansatz}(t)=0.
\end{align}
Waveform amplitudes and frequencies are reconstructed from the learned modulation functions (Appendix~\ref{sec:app_surrogate}), while the waveform phase is obtained by integrating the instantaneous frequency. Because small frequency errors accumulate into large phase errors, the agent introduces a simple low-order phase correction that removes coherent drift without absorbing the eccentric physical structure represented by the surrogate (Appendix~\ref{sec:app_bestfit_sur}).\\

\noindent \textbf{\textit{Simplifying and interpreting the learned equations: }}The surrogate is initially optimized for predictive accuracy without explicit constraints on expression complexity or physical organization. The resulting model contains 2955 polynomial--Fourier basis functions, making direct physical interpretation difficult. We therefore introduce a dedicated interpretation stage in which the agent reorganizes the learned expression into a compact hierarchy of physically meaningful components while preserving predictive accuracy.

Figure~\ref{fig:interpretation} summarizes this procedure. The agent first ranks basis functions according to their contribution to the prediction, then identifies recurring patterns in orbital harmonics, orbital-frequency scaling, spin dependence, and eccentricity. These structures are subsequently reorganized into a compact analytic representation that exposes the dominant physical dependencies while removing redundant complexity. Throughout this procedure, candidate simplifications are accepted only if they (approximately) preserve validation accuracy. Feature-importance analysis (Fig.~\ref{fig:interpretation}b) shows that the dominant learned corrections arise from sine-like orbital anomaly variations associated with eccentric orbital motion, while the remaining terms naturally separate into non-spin, symmetric-spin, and antisymmetric-spin sectors. We provide a detailed discussion of the best-fit surrogate in  Appendix~\ref{sec:app_bestfit_sur}.

The interpretation stage compresses the original 2955-term surrogate into the following compact symbolic representation:
\begin{align}
\delta \xi_{\rm amp}
\approx
\sum_{i}
\frac{e^{a_i}}{(1-e^2)^{1/2}}
\,x^{b_i}\,
\nu^{c_i}\,
\chiS^{d_i}\,
\chiA^{f_i}
\nonumber\\
\times
\left[
c_0^{(i)}
+
\sum_{k=1}^{3}
\alpha_k^{(i)}\cos(k\zeta)
+
\beta_k^{(i)}\sin(k\zeta)
\right].
\label{eq:resummed_formula}
\end{align}
where the complete coefficients are provided in the Appendix~\ref{sec:app_interp}. The compact representation preserves the predictive accuracy of the original polynomial model, achieving a median waveform mismatch of $\sim6.9\times10^{-4}$ and $R^2\simeq0.996$ for both waveform amplitude and frequency modulations.

Interestingly, the common factor $(1-e^2)^{-1/2}$ was not imposed during model construction, but emerges naturally during the interpretation stage as an effective resummation over the training domain ($e\le0.5$). Likewise, the leading corrections organize into combinations such as $e^2x\nu\sin\zeta$, $ex^2\chi_S\sin\zeta$, and $ex^2\chi_A\sin\zeta$, with remarkably consistent relative amplitudes between the amplitude and frequency surrogates. Rather than functioning solely as a numerical interpolant, the final surrogate provides an interpretable hierarchy of physical effects governing eccentric waveform corrections.\\

\noindent \textbf{\textit{Quantifying model accuracy: }}The compact surrogate closely reproduces the reference simulations throughout the binary evolution. The maximum cumulative phase error remains below $\sim0.2$ rad, substantially smaller than for the baseline model. We quantify waveform accuracy using the standard Advanced LIGO match, defined as the noise-weighted overlap between the surrogate and reference waveforms~\cite{Cutler:1994ys},
\begin{equation}
\begin{split}
\mathcal{M}_\mathrm{LIGO}
&\equiv
\langle
\tilde{h}_\mathrm{surrogate}
\mid
\tilde{h}_\mathrm{simulation}
\rangle
\\
&=
4\,\mathrm{Re}
\int_{f_{\min}}^{f_{\max}}
\frac{
\tilde{h}_\mathrm{surrogate}(f)
\tilde{h}_\mathrm{simulation}^{*}(f)
}
{S_\mathrm{LIGO}(f)}
\,df,
\end{split}
\label{eq:mismatch_definition}
\end{equation}
where $\tilde{h}(f)$ denotes the normalized Fourier-domain waveform satisfying $\langle\tilde{h}|\tilde{h}\rangle=1$, and $S_\mathrm{LIGO}(f)$ is the one-sided Advanced LIGO noise power spectral density~\cite{KAGRA:2013rdx}. The corresponding waveform mismatch is
\[
\mathcal{MM}_{\rm LIGO}
=
1-\mathcal{M}_{\rm LIGO}.
\]
More details are provided in Appendix~\ref{sec:app_metrics}.

Throughout the optimization, candidate implementations are evaluated against quantitative accuracy and runtime targets. The optimized orbital evolution reproduces the reference simulator to within $\lesssim1\%$ relative error, while successive surrogate refinements progressively reduce waveform mismatch and cumulative phase error (Fig.~\ref{fig:workflow}b; Appendix, Fig.~\ref{fig:supp_dynamics_progress}). The final implementation achieves an overall speedup of $\sim3.5\times$, including an $\sim8.4\times$ speedup in waveform evaluation ($\sim13$ ms per waveform), while maintaining sub-radian phase accuracy.

\begin{figure*}
	\centering
	\includegraphics[width=\linewidth]{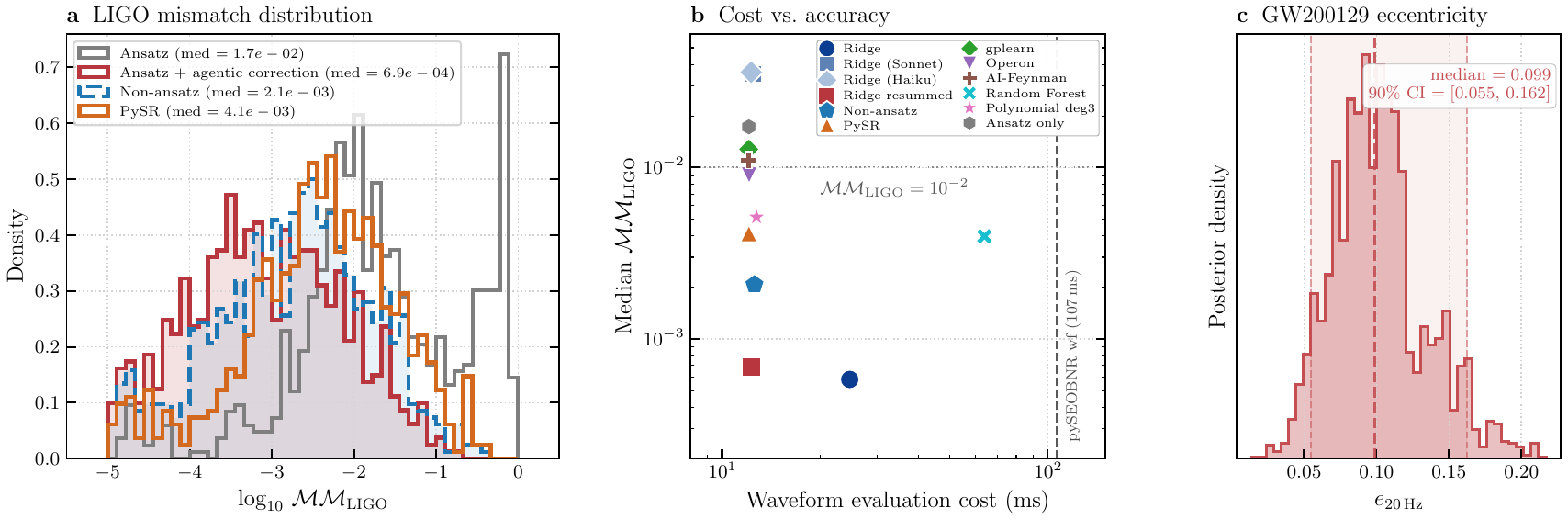}
	\caption{\textbf{Comparison of agent-built surrogate models with other approaches.}
    (a) Distribution of Advanced LIGO waveform errors $\mathcal{MM}_{\mathrm{LIGO}}$ [$\downarrow$ better, see Eq.~(\ref{eq:mismatch_definition})] across the validation set of binary properties. Agent-derived surrogates achieve substantially lower errors than baseline physics-informed and symbolic-regression methods, with median error $\sim 6.9\times10^{-4}$, compared to $\sim 10^{-2}$ for baseline models.
    (b) Trade-off between accuracy and computational cost. Points show median mismatch versus waveform evaluation time. Agent-built surrogates (which use Ridge regression) define a favorable accuracy--cost frontier, achieving $\sim 6.9\times10^{-4}$ accuracy at $\mathcal{O}(10)$ ms cost, outperforming symbolic-regression (PySR, gplearn, Operon, and AI-Feynman) and classical machine learning approaches. Dashed lines indicate the baseline simulation cost of \texttt{pySEOBNR}. 
    (c) \textbf{Eccentric binary black hole interpretation of GW200129.}
	Posterior distribution for the eccentricity at 20 Hz ($e_{20\mathrm{Hz}}$) inferred using the \GWAgent{} surrogate. The posterior supports moderate eccentricity, with median and 90\% credible interval $e_{20\mathrm{Hz}}=0.099^{+0.063}_{-0.044}$.}
    \label{fig:pareto}
\end{figure*}

The validation set is held out from all model fitting and model-selection steps. Performance is evaluated using waveform mismatch together with instantaneous-frequency and cumulative-phase errors. Across the validation set, the final surrogate achieves a median waveform mismatch of $\sim6.9\times10^{-4}$, approximately an order of magnitude lower than the baseline physics-informed model (Fig.~\ref{fig:pareto}; Appendix, Fig.~\ref{fig:mismatch_hist}). This level of agreement is well within the accuracy required for current GW data-analysis applications.\\

\begin{figure*}
    \centering
    \includegraphics[width=0.85\linewidth]{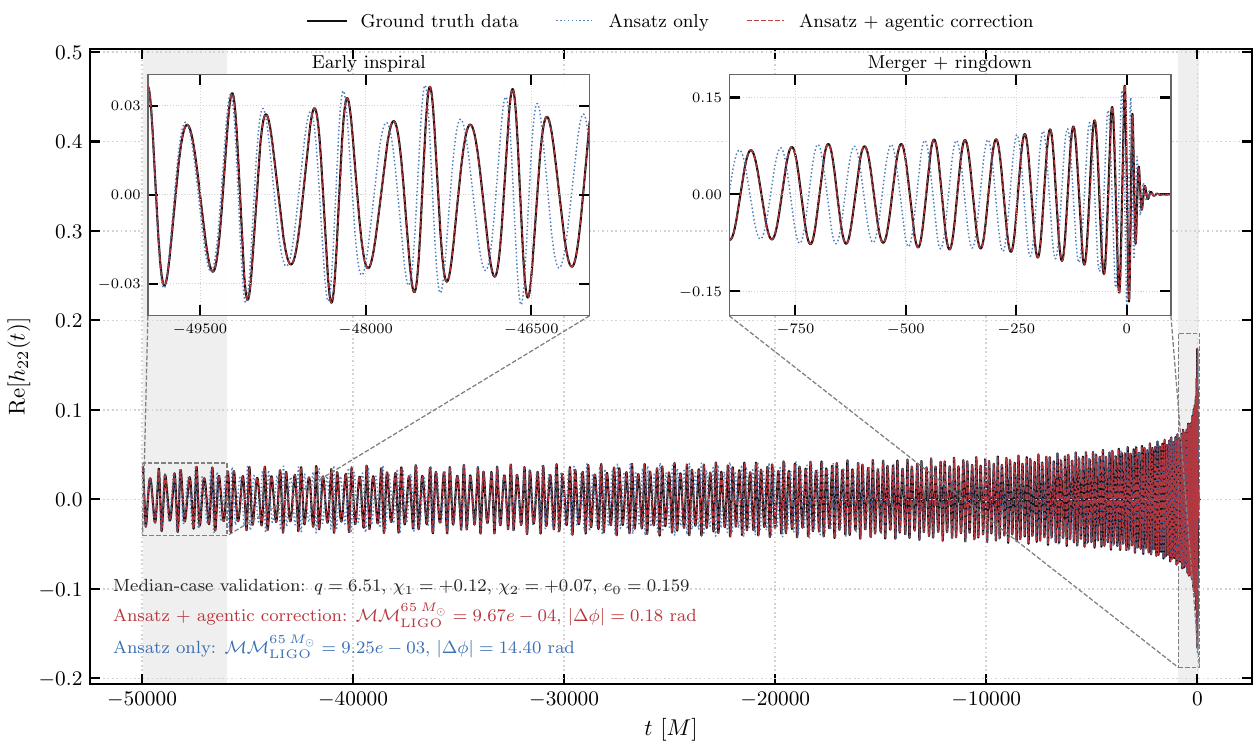}
    \caption{\textbf{Waveform reconstruction for a representative validation case.}
    Comparison of the waveform from numerical simulation data for an eccentric black hole binary (\texttt{pySEOBNR}, black), the analytic ansatz we use as a baseline (blue), and the agent-derived surrogate (red) for the binary parameters listed in the plot. Our surrogate model accurately tracks the full signal from inspiral through merger and final relaxation, achieving a mismatch $\mathcal{MM}_{\mathrm{LIGO}} \sim 6.9\times10^{-4}$ and maximum phase error $\sim 0.2$ rad.}
    \label{fig:s5_waveform_validation}
\end{figure*}

\noindent \textbf{\textit{Astrophysical inference for GW200129: }}To demonstrate its utility for scientific inference, we apply the surrogate to Bayesian parameter estimation of the GW event GW200129~\cite{KAGRA:2021vkt} using public LIGO--Virgo--KAGRA data and a standard compact-binary likelihood implemented in \texttt{bilby}~\cite{Ashton:2018jfp} (Appendix~\ref{sec:app_pe}). The inferred eccentricity posterior is shown in the right panel of Fig.~\ref{fig:pareto}, while the full posterior distributions are provided in the Appendix (Fig.~\ref{fig:supp_gw200129_corner}). To validate the surrogate in the inference setting, we compare approximately $2\times10^4$ likelihood evaluations against the reference simulations. The resulting log-likelihood differences are sharply peaked near zero, with median $\Delta\ln\mathcal{L}\approx-1.17$, corresponding to a signal-to-noise-ratio difference of approximately $0.4$. The inferred eccentricity is
\[
e_{20\,\mathrm{Hz}}
=
0.099^{+0.063}_{-0.044},
\]
obtained by evolving the sampled initial eccentricities using the same EOB dynamics employed for waveform generation.
The remaining source parameters are consistent with previous eccentric analyses within uncertainties~\cite{Gupte:2024jfe,Planas:2025jny,Tang:2026jvl}. Although the present surrogate models only the dominant quadrupolar mode $h_{22}(t)$, previous studies have shown that higher-order modes have little impact on the inferred eccentricity for this event~\cite{Tang:2026jvl}.
This application also provides a more stringent test than the original validation set; during parameter estimation, the surrogate is evaluated at parameter combinations selected by the data and the prior rather than by the training or validation design. We see that the log-likelihood differences remain small across the posterior samples, demonstrating that the surrogate generalizes beyond its original validation configurations and supports end-to-end GW inference.\\

\noindent \textbf{\textit{Comparison with alternative methods: }}We benchmark \GWAgent{} against conventional ML and symbolic-regression methods trained on the same residual-learning problem. The symbolic-regression baselines include \pySR{}~\cite{Cranmer2023InterpretableML}, \textcolor{linkcolor}{\texttt{gplearn}}~\cite{gplearn043}, \textcolor{linkcolor}{\texttt{Operon}}~\cite{Burlacu2020OperonCA}, and \textcolor{linkcolor}{\texttt{AI-Feynman}}~\cite{Udrescu2019AIFA}, each applied in a one-shot setting. The center panel of figure~\ref{fig:pareto} summarizes the resulting accuracy--cost trade-offs. Most baselines achieve waveform mismatches $\mathcal{MM}_{\rm LIGO}>10^{-3}$, whereas the agent-derived surrogate lies on the best accuracy--cost Pareto frontier. At evaluation times below $\sim12$ ms, waveform generation becomes dominated by phase integration rather than residual evaluation, producing the observed saturation in computational cost.
Among the symbolic-regression baselines, \pySR{} produces the most accurate expressions. A representative model for the waveform-frequency residual is
\begin{align}
\delta\xi_{\omega}^{\rm PySR}
&= e^2 \Bigg[
e\bigl(-\cos\zeta-\sin^2\zeta-0.079163484\bigr)-0.8009563
\notag\\
&\qquad
+\left(e-\frac{3.1693237\,x}{e}\right)
x\sin(2\zeta)
\left(\frac{-4.4655943}{\cos\zeta}\right)
\Bigg].
\end{align}
Although this expression achieves reasonable pointwise accuracy, it combines polynomial, trigonometric, and rational functions through nested terms such as $x/e$ and $1/\cos\zeta$, making physical interpretation difficult. By contrast, the agent-derived surrogate (Eq.~\ref{eq:resummed_formula}) separates eccentricity, orbital harmonics, orbital-frequency scaling, and spin dependence into physically meaningful components while achieving higher waveform accuracy. The corresponding \pySR{} amplitude model and additional symbolic-regression results are provided in the Appendix~\ref{sec:app_benchmark}.

We further compare against classical ML baselines, including random forests~\cite{scikit-learn} and low-order polynomial regression in Figure~\ref{fig:pareto}. Random forests achieve low training error but generalize less robustly while providing little physical interpretability, whereas polynomial models remain interpretable but cannot accurately capture the time-dependent eccentric waveform structure. These comparisons illustrate that the contribution of \GWAgent{} is not simply selecting a regression algorithm. Instead, the workflow jointly optimizes the physical decomposition, surrogate representation, basis construction, and post-fit interpretation under continuous simulation-based validation, producing compact analytic models that remain both accurate and physically interpretable.\\

\noindent \textbf{\textit{Discussion: }}Our results demonstrate a natural role for language-model agents in areas of scientific modeling where candidate solutions can be executed, quantitatively evaluated, and iteratively refined against high-fidelity simulations. In this setting, the language model does not replace the simulator or the underlying physical theory. Instead, it explores scientific representations while the simulator provides a continuous grounding signal through objective validation metrics. We note that search space explored by \GWAgent{} is substantially broader than the algebraic-expression space considered by conventional symbolic-regression algorithms, encompassing physical decompositions, residual formulations, basis construction, implementation strategies, and symbolic interpretation.

The GW application provides a particularly demanding demonstration of this framework because the surrogate must simultaneously preserve eccentric waveform structure, maintain phase coherence over long signals, and remain sufficiently efficient for large-scale Bayesian inference. The resulting surrogate combines human-specified physical priors with agent-discovered corrections, illustrating that domain expertise and agentic optimization are complementary rather than competing approaches. Equally important, the interpretation stage transforms a high-dimensional numerical surrogate into a compact symbolic representation, showing that agentic workflows can produce models that are not only accurate but also scientifically interpretable.

The present study also illustrates several limitations. The surrogate is trained and validated over a finite parameter domain, uses the effective-one body simulations rather than the more accurate numerical relativity simulations as the reference model, and includes only the dominant waveform mode in the GW200129 analysis. These choices are appropriate for demonstrating the methodology, but they should not be interpreted as limitations of the framework itself. Extending the workflow to higher-order modes, spin precession, broader eccentricity ranges, and larger numerical relativity catalogs represents a natural direction for future work.

The development process also highlights the importance of rigorous validation. Agentic workflows can produce implementations that appear plausible yet fail scientific correctness tests through validation leakage, misleading runtime optimizations, or incorrect physical reconstruction. Representative examples are provided in the Appendix~\ref{sec:app_failure_modes}. These failure modes motivate the simulation-grounded design adopted here and demonstrate that quantitative validation remains essential for trustworthy scientific AI. Human expertise remains indispensable for defining scientific objectives, designing validation metrics, and identifying physically inconsistent solutions.

Although demonstrated here for GW surrogate modeling, the underlying methodology extends naturally to other simulation-driven scientific problems. Candidate applications include waveform modeling with spin precession, cosmological hydrodynamical simulations, weather prediction, fluid dynamics, molecular simulations, and computational chemistry~\cite{lam2023graphcast,Bi2023AccurateMG,Schtt2017SchNetA,Li2020FourierNO,Raissi2019PhysicsinformedNN,Mancini2021COSMOPOWEREC}. More broadly, our results suggest that simulation-grounded language-model agents offer a practical route toward discovering accurate, efficient, and interpretable scientific models, complementing conventional machine learning wherever quantitative validation against high-fidelity simulations is possible. In the specific setting of symbolic modeling, the agentic workflow moves beyond a complementary role: it surpasses one-shot symbolic-regression baselines on the high-accuracy problems studied here, pointing toward a new route to data-driven symbolic equation discovery~\cite{Shojaee2024LLMSRSE,Yang2026ThinkLA, Grayeli2024SymbolicRW, su2026strideselfreflectiveagentframework, xia2025sr, wang2025drsr, guo2025sr, pang2026deliberateevolutionagenticreasoning}.

\noindent \textbf{\textit{Code availability: }}The \texttt{GWAgent} framework is publicly available at \url{https://github.com/tousifislam/GWAgent}. The repository includes the structured workflow protocol used to initialize the agent, the generated datasets, trained surrogate models, evaluation outputs, and scripts required to reproduce the results presented in this work. The GW200129 analysis uses publicly available LIGO--Virgo--KAGRA strain data.

\begin{acknowledgments}
We thank Scott Field, Gaurav Khanna, Siddharth Mishra-Sharma, Mark Ho-Yeuk Cheung, Niels Deppe, Keefe Mitman, and Peter James Nee for helpful discussions.
T.I. is supported in part by the National Science Foundation under Grant No.~PHY-2309135 and the Gordon and Betty Moore Foundation Grant No.~GBMF7392. Computational resources were provided by facilities purchased with funding from the National Science Foundation (CNS-1725797) and administered by the Center for Scientific Computing (CSC). The CSC is supported by the California NanoSystems Institute and the Materials Research Science and Engineering Center (MRSEC; NSF DMR 2308708) at UC Santa Barbara.
\end{acknowledgments}

\bibliography{references}

\appendix

\begin{figure}
    \centering
    \includegraphics[width=\columnwidth]{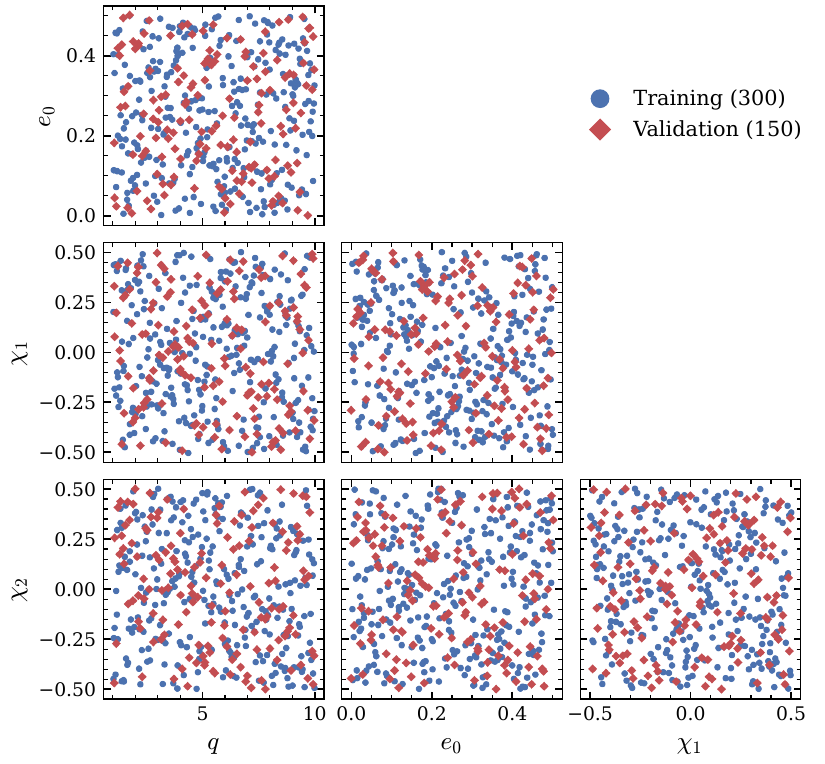}
    \caption{\textbf{Sampling of the parameter space for surrogate construction.}
    Pairwise distributions and marginal histograms for the intrinsic parameters $(e_0, q, \chi_1, \chi_2)$ used in training and validation. Blue points (and histograms) correspond to the training set (300 configurations), while red points denote the validation set (150 configurations). The samples are approximately uniformly distributed across the specified parameter ranges, with no significant gaps or clustering, ensuring adequate coverage for surrogate training and reliable assessment of generalization performance. More details are in Appendix~\ref{sec:app_param}.}
    \label{fig:supp_parameter_space}
\end{figure}

\section{Waveform setup and parameter sampling}
\label{sec:app_param}
The surrogate is constructed from eccentric, spinning BBH waveforms generated with the EOB model \texttt{SEOBNRv5EHM} implemented in \texttt{pySEOBNR} package~\cite{Gamboa:2024hli,Gamboa:2024imd}. Throughout this work, the EOB model serves as the reference simulator. We model the dominant $(\ell,m)=(2,2)$ GW mode over the parameter ranges
$q\in[1,10]$, $|\chi_{1,2}|\le0.5$, $e_0\in[0,0.5]$, and $\zeta_0\in[0,2\pi]$. Depending on the binary configuration, the waveforms span durations of approximately $10^3$--$10^5\,M$.

Training and validation configurations are generated independently using Latin hypercube sampling over the four-dimensional parameter space $(q,\chi_1,\chi_2,e_0)$ with \textcolor{linkcolor}{\texttt{scipy.stats.qmc.LatinHypercube}}. For each binary, the orbital anomaly $\zeta_0$ is drawn independently from a uniform distribution on $[0,2\pi]$. The training and validation sets contain 300 and 150 waveforms, respectively, and the resulting parameter coverage is shown in Fig.~\ref{fig:supp_parameter_space}. The validation set is excluded from surrogate fitting, hyperparameter optimization, and model selection, and is used exclusively for the mismatch, phase-error, likelihood, and robustness studies.

Because relatively few spinning eccentric NR simulations are currently available, we use EOB waveforms for surrogate construction. The workflow itself is independent of the underlying waveform generator and can be applied directly to NR simulations as larger eccentric catalogs become available.

\section{Agentic implementation}
\label{sec:app_agentic}
The workflow described in the main text is implemented within the open-source \GWAgent{} framework, which coordinates code generation, execution, validation, and iterative refinement while interfacing with the reference waveform simulator and the surrogate-construction pipeline. Intermediate products, including fitted models, generated source code, diagnostic plots, validation metrics, and optimization logs, are produced automatically and stored throughout the optimization process.

The workflow is initialized from a structured Markdown protocol rather than a single free-form prompt. The protocol specifies the scientific objective, available simulation tools, staged tasks, validation metrics, target runtimes, stopping criteria, and expected output artifacts. This protocol remains fixed throughout an optimization run and defines the scientific specification independently of the underlying language model. Human oversight enters by defining the scientific objectives, validation criteria, and acceptance thresholds, while the language model is responsible for proposing, implementing, executing, and refining candidate solutions within these constraints (Table~\ref{tab:agent_human_roles}).

The implementation follows a \textcolor{linkcolor}{Ralph-loop}-like execution framework~\footnote{\href{https://github.com/anthropics/claude-code/tree/main/plugins/ralph-wiggum}{https://github.com/anthropics/claude-code/tree/main/plugins/ralph-wiggum}}. At each iteration, the agent proposes a modification to the dynamical evolution, surrogate model, or analysis pipeline, executes the modified implementation, evaluates objective scientific metrics including runtime, waveform mismatch, and cumulative dephasing, and retains the modification only if it improves the current solution or satisfies the prescribed validation criteria. Successful and unsuccessful attempts are preserved as context for subsequent iterations, allowing the workflow to progressively refine both accuracy and computational efficiency through quantitative feedback.

The primary implementation uses Claude Opus 4.6 through the Claude Code interface, which provides sandboxed terminal access, file I/O, and iterative code execution. We also evaluated Claude Sonnet 4.6 (\texttt{claude-sonnet-4-6}) and Claude Haiku 4.5 (\texttt{claude-haiku-4-5-20251001}) under the same workflow protocol and validation criteria. All models support a 200k-token context window and were executed with temperature zero for reproducibility. The workflow is model-agnostic, allowing different frontier coding agents to be substituted without modifying the scientific protocol.

Candidate implementations are executed with direct access to the reference waveform simulator, numerical libraries, and intermediate analysis products. The implementation is written in Python and uses NumPy, SciPy, Numba, and \texttt{pySEOBNR}. Intermediate models are serialized automatically, while waveform validation, mismatch calculations, and diagnostic figures are generated programmatically. The complete workflow protocol, prompts, generated datasets, trained surrogate models, evaluation outputs, and source code are distributed through the public \GWAgent{} repository.

All agentic runs were performed on a MacBook Pro equipped with an Apple M3 Pro processor (11 cores, 36\,GB unified memory) running macOS~26.2.

\section{Surrogate model formulation}
\label{sec:app_surrogate}
This section presents the modulation variables, analytic baseline, residual formulation, and waveform reconstruction used throughout the surrogate model.

Eccentric effects are quantified relative to the corresponding quasi-circular waveform. For a waveform mode $h_{\ell m}(t;\boldsymbol{\lambda})$ and its quasi-circular counterpart $h_{\ell m}(t;\boldsymbol{\lambda}^0)$, we define the amplitude and frequency modulation functions~\cite{Islam:2024rhm,Islam:2024bza,Islam:2024zqo}
\begin{align}
\xi_{\ell m}^{A}(t)
&=
\frac{2}{\ell}
\frac{
A_{\ell m}(t;\boldsymbol{\lambda})
-
A_{\ell m}(t;\boldsymbol{\lambda}^0)
}
{
A_{\ell m}(t;\boldsymbol{\lambda}^0)
},
\\
\xi_{\ell m}^{\omega}(t)
&=
\frac{
\omega_{\ell m}(t;\boldsymbol{\lambda})
-
\omega_{\ell m}(t;\boldsymbol{\lambda}^0)
}
{
\omega_{\ell m}(t;\boldsymbol{\lambda}^0)
}.
\end{align}
These modulation functions are approximately universal across different spherical-harmonic modes, allowing eccentric effects to be represented by a small set of functions. Throughout this work we model only the dominant $(2,2)$ mode,
\begin{align}
\xi_{\rm amp}(t)
&\equiv
\xi_{22}^{A}(t),
\qquad
\xi_{\omega}(t)
\equiv
\xi_{22}^{\omega}(t).
\end{align}

To capture the leading eccentric modulation, we introduce the physics-informed analytic baseline~\cite{Gamboa:2024hli}
\begin{equation}
f_{\rm ansatz}(t)
=
\left|
f_{\rm lead}(e,\zeta)
+
f_{\rm PN}(e,\zeta,x,\nu)
\right|
-
1,
\label{eq:ansatz_expression}
\end{equation}
where $\nu=q/(1+q)^2$ is the symmetric mass ratio,
\begin{align}
f_{\rm lead}(e,\zeta)
&=
\frac{
4
+
2e^2e^{2i\zeta}
+
ee^{-i\zeta}
+
5ee^{i\zeta}
}
{
4(1-e^2)
},
\\
f_{\rm PN}(e,\zeta,x,\nu)
&=
\frac{
xe
}
{
(1-e^2)^2
}
\,
\mathcal{C}(e,\zeta,\nu),
\end{align}
and $\mathcal{C}(e,\zeta,\nu)$ contains the higher-order harmonic contributions. This ansatz captures the dominant oscillatory structure associated with eccentric orbital motion.

The surrogate is trained to model only the residual corrections beyond this analytic baseline,
\begin{align}
\xi_{\rm amp}(t)
&=
f_{\rm ansatz}(t)
+
\delta\xi_{\rm amp}(t),
\\
\xi_{\omega}(t)
&=
f_{\rm ansatz}(t)
+
\delta\xi_{\omega}(t),
\end{align}
where the residuals are modeled as functions of the evolving orbital variables $(e(t),x(t),\zeta(t))$ together with the intrinsic binary parameters $(q,\chi_1,\chi_2)$.

The eccentric waveform is reconstructed from the quasi-circular waveform using the learned modulation functions,
\begin{align}
A_{\ell m}(t)
&=
A_{\ell m}^{\rm circ}(t)
\left[
1+\frac{\ell}{2}\xi_{\rm amp}(t)
\right],
\\
\omega_{\ell m}(t)
&=
\omega_{\ell m}^{\rm circ}(t)
\left[
1+\xi_{\omega}(t)
\right],
\end{align}
with the phase obtained by integrating the instantaneous frequency,
\begin{align}
\phi_{\ell m}(t)
=
\int
\omega_{\ell m}(t)\,
dt.
\end{align}
Residual frequency errors accumulate under integration and produce a slowly varying phase drift. This effect is corrected using the low-order phase adjustment described in Appendix~\ref{sec:app_bestfit_sur}.

\begin{table}[t]
\centering
\begin{tabular}{ll}
\hline
\textbf{Workflow component} & \textbf{Responsibility} \\
\hline
Scientific objective        & Human \\
Reference simulator         & Human \\
Validation metrics          & Human \\
Physics-informed ansatz     & Human \\
Residual formulation        & Human + Agent \\
Feature basis               & Agent \\
Regression model            & Agent \\
Hyperparameter tuning       & Agent \\
Model simplification        & Agent \\
Physical interpretation     & Agent \\
Symbolic compression        & Agent \\
\hline
\end{tabular}
\caption{Division of responsibilities between human guidance and agent-driven optimization within the \GWAgent{} workflow. Human input specifies the scientific objective, physical priors, and validation protocol, while the agent discovers the surrogate representation and compact symbolic model through iterative simulation-grounded optimization. More details are in Appendix~\ref{sec:app_agentic}.}
\label{tab:agent_human_roles}
\end{table}

\begin{figure*}
    \centering
    \includegraphics[width=0.8\textwidth]{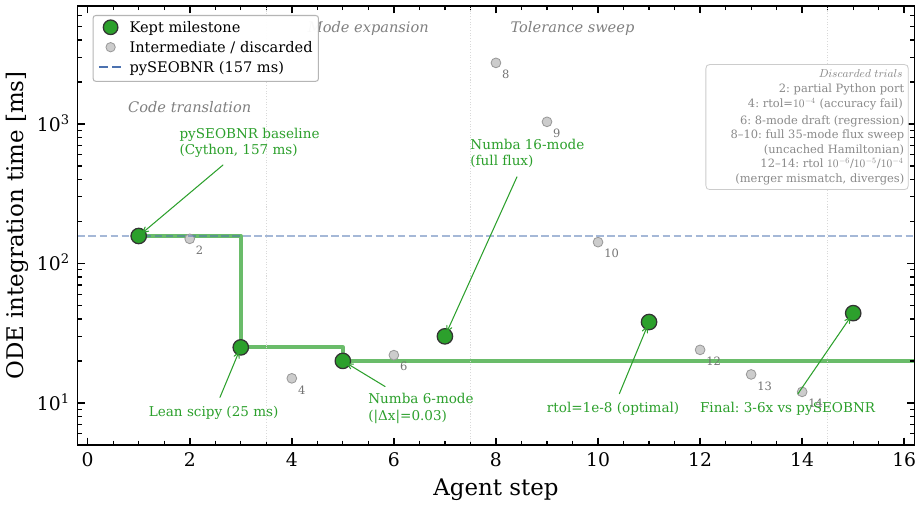}
    \caption{\textbf{Agent-driven optimization of the dynamical evolution.}
    The agent iteratively refines the implementation of the EOB equations of motion, reducing ODE integration time while maintaining accuracy. Green markers indicate retained improvements, while gray markers denote intermediate or discarded trials. The final implementation achieves a substantial reduction in computational cost relative to the \texttt{pySEOBNR} baseline. More details are in the main text.}
    \label{fig:supp_dynamics_progress}
\end{figure*}

\begin{figure}
    \centering
    \includegraphics[width=0.9\columnwidth]{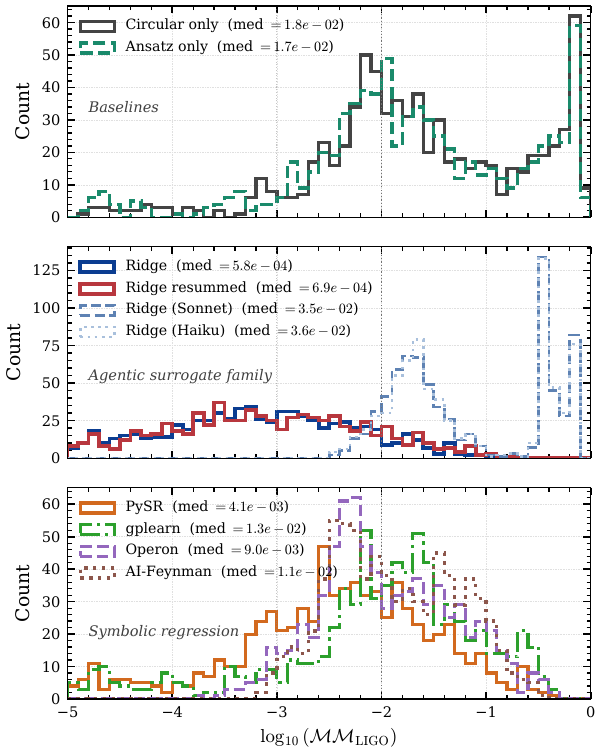}
    \caption{\textbf{Extended mismatch distributions across modeling approaches.}
    Histograms of Advanced LIGO mismatch $\mathcal{MM}$ (shown in $\log_{10}$ scale) evaluated over the validation set. 
    \textbf{Top:} Baseline models, including circular and ansatz-only waveforms, exhibit mismatches at the $\sim 10^{-2}$ level. 
    \textbf{Middle:} Agentic surrogate models produce sharply peaked distributions at low mismatch values, with median $\sim 6.9\times10^{-4}$, while reduced-capacity variants show degraded performance. 
    \textbf{Bottom:} symbolic-regression methods yield broader distributions with significant tails extending to $\gtrsim 10^{-3}$--$10^{-2}$. 
    These results demonstrate that the agent-derived surrogate achieves both higher accuracy and substantially improved robustness compared to baseline and symbolic approaches. More details are in the main text and in Appendix~\ref{sec:app_benchmark}.}
    \label{fig:mismatch_hist}
\end{figure}

\section{Best-fit agentic surrogate models}
\label{sec:app_bestfit_sur}
This section describes the surrogate model selected by the agent together with the feature basis, phase correction, and computational performance.

Feature-importance analysis of the residuals $\delta\xi_{\rm amp}$ and $\delta\xi_{\omega}$ shows that eccentricity is the dominant parameter, accounting for approximately $80$--$85\%$ of the total variance. The remaining structure is primarily associated with orbital harmonics, the dimensionless orbital frequency $x$, and weaker spin dependence through $\chi_S$ and $\chi_A$. This naturally motivates a polynomial--Fourier basis of the form
\begin{align}
\Phi_{abcdf,k}(t)
=
e^a x^b \nu^c \chi_S^d \chi_A^f
\begin{cases}
1, & k=0,\\
\cos(k\zeta),\;\sin(k\zeta), & k>0,
\end{cases}
\end{align}
which captures both the secular parameter dependence and the oscillatory eccentric modulation.

Among the models explored by the agent, the best-performing surrogate is a Ridge-regression model constructed from this basis. The final configuration uses polynomial orders
\begin{align}
0\le a\le5,\qquad
0\le b\le3,\qquad
0\le d,f\le1,\qquad
0\le k\le7,
\end{align}
with regularization parameter $\alpha=10^{-6}$, resulting in 2955 basis functions. The corresponding surrogate is
\begin{align}
\xi_{\rm amp}^{\rm sur}(t)
&=
f_{\rm ansatz}(t)
+
\sum_{i=1}^{2955}
c_i^{(A)}
\Phi_i(e,x,\zeta;q,\chi_{1,2}),
\\
\xi_{\omega}^{\rm sur}(t)
&=
f_{\rm ansatz}(t)
+
\sum_{i=1}^{2955}
c_i^{(\omega)}
\Phi_i(e,x,\zeta;q,\chi_{1,2}),
\end{align}
where $\Phi_i$ denotes the basis functions defined above. Prior to phase correction, the surrogate achieves a median waveform mismatch of approximately $10^{-3}$ while generalizing robustly across the validation set.

Residual errors in $\xi_{\omega}(t)$ accumulate under time integration and produce a slowly varying phase drift. We therefore introduce a low-order polynomial correction,
\begin{align}
\Delta\phi_{\rm corr}(t)
=
\sum_{j=1}^{5}
c_jT^j,
\qquad
T=
\frac{t-t_0}{t_{\rm end}-t_0},
\end{align}
where the coefficients $c_j$ are obtained by least-squares fitting the residual phase difference. Applying this correction reduces the median waveform mismatch to approximately $6.9\times10^{-4}$ while limiting the cumulative phase error to $\lesssim0.2$ rad.

The optimized implementation evaluates a waveform in approximately $13$ ms, corresponding to an $\sim8.4\times$ speedup relative to the reference implementation. Including the optimized orbital evolution, the complete workflow achieves an overall speedup of approximately $3.5\times$.

\section{Accuracy metrics}
\label{sec:app_metrics}
Waveform accuracy is quantified using the Advanced LIGO mismatch defined in the main text. Before computing the overlap, the surrogate and reference waveforms are maximized over relative time and phase shifts $(t_c,\phi_c)$. The mismatch is evaluated using the Advanced LIGO design sensitivity curve~\cite{KAGRA:2013rdx} at five detector-frame total masses uniformly spaced between $20\,M_\odot$ and $200\,M_\odot$.

Phase accuracy is quantified using the maximum cumulative phase error,
\begin{align}
\Delta\phi_{\rm LIGO}
=
\max_t
\left|
\phi_{\rm surrogate}(t)
-
\phi_{\rm simulation}(t)
\right|,
\end{align}
computed after waveform alignment.

\section{Interpreting the learned surrogate}
\label{sec:app_interp}
This section describes the physical structure identified in the fitted surrogate and the procedure used to compress the full polynomial--Fourier expansion into the compact symbolic representation presented in the main text.

The residual surrogate can be written as
\begin{align}
\delta\xi(t)
=
\sum_i
c_i\,
\Phi_i(e,x,\zeta;q,\chi_1,\chi_2),
\end{align}
where $\Phi_i$ denotes the polynomial--Fourier basis introduced in Appendix~\ref{sec:app_bestfit_sur}. The fitted model contains 2955 basis functions. The harmonic structure of the residual is well described by orbital harmonics up to $k_{\rm max}=6$, while contributions beyond this order are negligible. Truncating the expansion to $k\le3$ preserves the surrogate accuracy while reducing the effective model size from 2955 to 1379 basis functions. After the phase correction described in Appendix~\ref{sec:app_bestfit_sur}, this reduced representation achieves a median validation mismatch of approximately $6.9\times10^{-4}$.

\begin{table*}[t]
\centering
\caption{\textbf{Leading coefficients of the learned surrogate.}
The ten largest coefficients in the amplitude residual model. The corresponding coefficients in the frequency residual model exhibit similar structure, differing primarily by an overall rescaling. The final column indicates the approximate physical interpretation of each term. More details are in Appendix~\ref{sec:app_interp}.}
\label{tab:top10}
\small
\begin{tabular}{clrrl}
\hline
Rank & Basis function & $c_{\rm amp}$ & $c_{\omega}$ & Interpretation \\
\hline
1  & $e^2x\nu\sin\zeta$                    & $+100.7$ & $+135.5$ & 1PN $\times$ Im$(h_{22}^{\rm ecc})$ \\
2  & $ex^2\chi_S\sin\zeta$                 & $+90.6$  & $+139.8$ & Spin-orbit (1.5PN) \\
3  & $ex\nu\chi_S\sin\zeta$                & $+79.5$  & $+113.3$ & Spin-orbit (1.5PN) \\
4  & $ex^2\chi_A\sin\zeta$                 & $+70.9$  & $+110.8$ & Spin-orbit ($\Delta$) \\
5  & $ex\nu\chi_A\sin\zeta$                & $+69.9$  & $+99.4$  & Spin-orbit ($\Delta$) \\
6  & $e^2x^2\cos\zeta$                     & $+61.3$  & $+69.1$  & 2PN correction \\
7  & $e^2x\chi_S\sin\zeta$                 & $+60.2$  & $+86.9$  & Spin-orbit, $e^2$ \\
8  & $e^3x\nu\sin\zeta$                    & $+58.0$  & $+79.1$  & Higher-order eccentricity \\
9  & $e^2x\nu\cos\zeta$                    & $-57.0$  & $-75.9$  & 1PN Re$(h_{22}^{\rm ecc})$ \\
10 & $e^2x\chi_A\sin\zeta$                 & $+53.3$  & $+77.8$  & Spin-orbit ($\Delta,e^2$) \\
\hline
\end{tabular}
\end{table*}

At fixed harmonic order, the learned coefficients exhibit a clear perturbative hierarchy in eccentricity and orbital frequency,
\begin{align}
\{e,e^2,e^3,\ldots\},
\qquad
\{x,x^2,x^3,\ldots\},
\end{align}
consistent with the structure expected from PN expansions. Spin effects enter primarily through the symmetric and antisymmetric spin combinations,
\begin{align}
\chi_S
=
\frac{\chi_1+\chi_2}{2},
\qquad
\chi_A
=
\frac{\chi_1-\chi_2}{2},
\end{align}
and appear predominantly as corrections to the leading eccentric terms rather than as independent structures.

Ranking the fitted coefficients by magnitude reveals that the dominant corrections are governed by $\sin\zeta$ harmonics. This behavior arises naturally because the analytic baseline removes the leading eccentric modulation, leaving the residual to reconstruct the remaining phase-dependent structure. At leading order,
\begin{align}
{\rm Im}\!\left(h_{22}^{\rm ecc}\right)
\Big|_{\rm 0PN}
\supset
\frac{e\sin\zeta}{1-e^2},
\end{align}
which produces the characteristic $e^a\sin\zeta$ dependence recovered by the surrogate. Likewise, spin-orbit contributions appear with the expected PN scaling,
\begin{align}
\delta h_{22}^{\rm SO}
\sim
x^{3/2}
e
\left(
\chi_Sf_S(\nu)
+
\chi_A\Delta f_A(\nu)
\right),
\end{align}
although the polynomial basis represents the fractional power through nearby integer powers of $x$. The dominant coefficients together with their PN interpretation are listed in Table~\ref{tab:top10}.

Grouping the leading coefficients according to their physical origin produces the compact decomposition
\begin{equation}
\begin{aligned}
\delta\xi_{\rm amp}
&\approx
\underbrace{
\sum_{a=1}^{5}
\sum_{c=0}^{2}
c_{a,c}^{(\rm ns)}
e^a\nu^c
(A_xx+B_xx^2)
\sin\zeta
}_{\rm non\mbox{-}spin}
\\
&\quad+
\underbrace{
\sum_{a=1}^{4}
e^a
(\gamma_1^{(S)}x+\gamma_2^{(S)}x^2+\gamma_3^{(S)}x\nu)
\chi_S
\sin\zeta
}_{\rm symmetric\ spin}
\\
&\quad+
\underbrace{
\sum_{a=1}^{4}
e^a
(\gamma_1^{(A)}x+\gamma_2^{(A)}x^2+\gamma_3^{(A)}x\nu)
\chi_A
\sin\zeta
}_{\rm antisymmetric\ spin}
+\cdots,
\end{aligned}
\label{eq:compact_physical}
\end{equation}
where the omitted terms correspond to smaller harmonic corrections. The frequency surrogate exhibits the same functional structure, differing primarily through an overall rescaling of the coefficients by factors of approximately $1.3$--$1.4$.

Finally, we examine whether the polynomial eccentricity expansion can be represented by a compact resummed form proportional to $(1-e^2)^{-p}$. A systematic scan over $p$ consistently favors
\begin{align}
p=\frac12,
\end{align}
across all PN sectors. This exponent should be interpreted as an effective resummation over the training domain $e\le0.5$ rather than an exact theoretical prediction. Applying this resummation yields the compact symbolic surrogate presented in Eq.~(\ref{eq:resummed_formula}) of the main text.

\section{Benchmarking against machine learning and symbolic-regression}
\label{sec:app_benchmark}
We benchmark \GWAgent{} against conventional machine-learning and symbolic-regression methods, including \pySR{}~\cite{Cranmer2023InterpretableML}, \textcolor{linkcolor}{\texttt{gplearn}}~\cite{gplearn043}, \textcolor{linkcolor}{\texttt{Operon}}~\cite{Burlacu2020OperonCA}, and \textcolor{linkcolor}{\texttt{AI-Feynman}}~\cite{Udrescu2019AIFA}. All models are trained on the same dataset consisting of approximately $2\times10^4$ samples drawn from the 300 training waveforms using the residual modulation functions $\delta\xi_{\rm amp}$ and $\delta\xi_{\omega}$ defined in Appendix~\ref{sec:app_surrogate} as targets. The symbolic-regression methods are executed in a one-shot setting, whereas \GWAgent{} performs iterative optimization using simulation-based validation throughout the search.

Among the symbolic-regression baselines, \pySR{} produces the most competitive expressions. The best models identified for the waveform-frequency and amplitude residuals are
\begin{align}
\delta\xi_{\omega}^{\rm PySR}
&= e^2 \Bigg[
e\bigl(-\cos\zeta - \sin^2\zeta - 0.07916\bigr) - 0.80096
\notag\\
&\qquad
+ \left(e - \frac{3.16932\,x}{e}\right)
x\sin(2\zeta)
\left(\frac{-4.46559}{\cos\zeta}\right)
\Bigg],
\end{align}

\begin{align}
\delta\xi_{\rm amp}^{\rm PySR}
&=
\frac{e^2}{0.93157}
\Bigg[
x
\left(
\cos(2\zeta)
+
e^2
\left(
\frac{\cos\zeta/0.59483+0.59892}{-x}
\right)
\right)
\notag\\
&\qquad
+
x^2
\left(
\frac{13.87375(\cos\zeta+\sin\zeta)}{e}
\right)
-
0.62500
\Bigg].
\end{align}

These expressions achieve reasonable pointwise accuracy only at relatively high complexity ($\sim30$--$40$) and do not exhibit a clear separation into harmonic, PN, or spin-dependent structures. In particular, nested rational terms such as $x/e$ and $1/\cos\zeta$ obscure physical interpretation.

\textcolor{linkcolor}{\texttt{Operon}} produces expressions of comparable complexity and predictive accuracy, for example
\begin{align}
\delta\xi_{\omega}^{\rm Operon}
&=
-0.00078
+
0.9928
\Bigg[
e^2
\Bigg(
e(-0.12-\cos\zeta-\sin^2\zeta)
\notag\\
&\qquad
-0.78+
x
\left(
\frac{x\sin(2\zeta)}{\cos\zeta}
\left(
-3.8+0.3x-\frac{2.5x}{e}
\right)
\right)
\Bigg)
\Bigg],
\end{align}

\begin{align}
\delta\xi_{\rm amp}^{\rm Operon}
&=
0.0004
+
1.0057
\Bigg[
e^2
\Bigg(
x
\left(
\cos(2\zeta)
-
\frac{e^2(\cos\zeta+0.65)}{x}
\right)
\notag\\
&\qquad
+
x^2
\left(
\frac{\cos\zeta+\sin\zeta}{e}
\left(
1.4+\frac{x}{e+0.18}
\right)
\right)
-
0.69
\Bigg)
\Bigg].
\end{align}

The full symbolic expressions are provided with the supplementary repository. Both \pySR{} and \textcolor{linkcolor}{\texttt{Operon}} achieve validation scores of approximately $R^2_{\rm val}\sim0.95$--$0.96$, but require substantially more complicated symbolic representations than the final \GWAgent{} surrogate.

Figure~\ref{fig:mismatch_hist} compares the waveform mismatch distributions. Although symbolic-regression methods recover some of the dominant harmonic structure, they exhibit broader mismatch distributions with significantly larger high-error tails. The \GWAgent{} surrogate instead produces a sharply peaked distribution centered near $\mathcal{MM}_{\rm LIGO}\simeq6.9\times10^{-4}$, while the symbolic-regression baselines extend to mismatches of order $10^{-3}$--$10^{-2}$.

\begin{figure}
    \centering
    \includegraphics[width=\columnwidth]{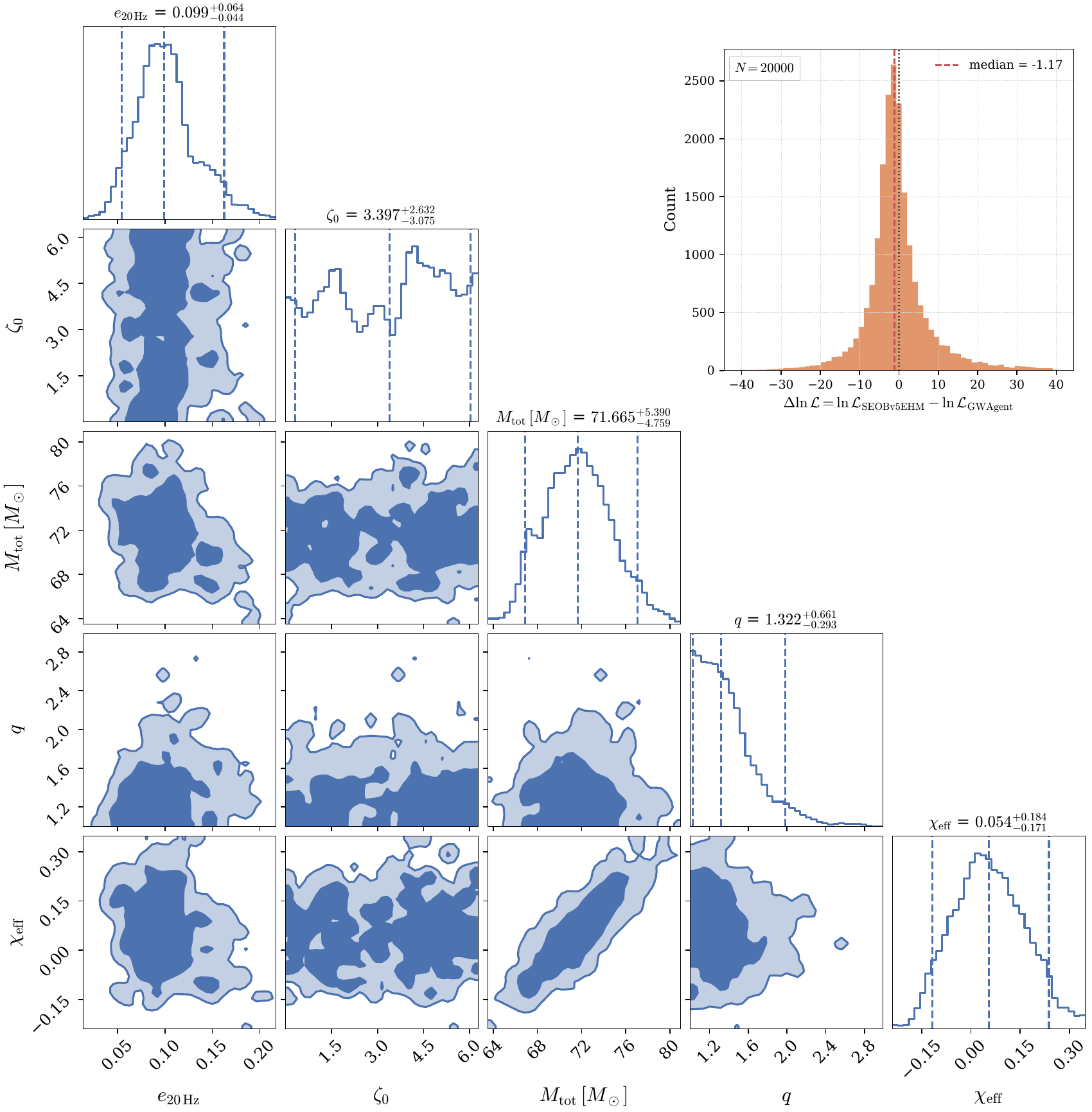}
    \caption{\textbf{Parameter estimation for GW200129 using the \texttt{GWAgent} surrogate.}
    \textbf{Left:} Posterior distributions for the eccentricity at 20 Hz ($e_{20\,\mathrm{Hz}}$), relativistic anomaly $\zeta_0$, total detector-frame mass $M_{\mathrm{tot}}$, mass ratio $q$, and effective inspiral spin $\chi_{\mathrm{eff}}$. Diagonal panels show marginalized distributions, while off-diagonal panels show joint constraints. Dashed lines indicate medians and 90\% credible intervals. The posterior supports a nonzero eccentricity at detector frequencies, $e_{20\,\mathrm{Hz}} \approx 0.10^{+0.06}_{-0.04}$.
    \textbf{Right:} Distribution of log-likelihood differences between \texttt{SEOBNRv5EHM} and \texttt{GWAgent}, $\Delta \ln \mathcal{L} = \ln \mathcal{L}_{\rm SEOBv5EHM} - \ln \mathcal{L}_{\rm GWAgent}$, evaluated over $\sim 2\times10^4$ samples. The distribution is sharply peaked near zero (median $\approx -1.17$), indicating close agreement between the surrogate and the reference model across the posterior support. More details are in Appendix~\ref{sec:app_pe}.}
    \label{fig:supp_gw200129_corner}
\end{figure}

\section{Parameter estimation for GW200129}
\label{sec:app_pe}

We perform Bayesian parameter estimation for the GW event GW200129~\cite{KAGRA:2021vkt} using the \texttt{bilby} inference framework~\cite{Ashton:2018jfp}. The analysis employs the standard compact-binary likelihood together with publicly available LIGO--Virgo--KAGRA strain data and power spectral densities~\cite{GWTC3_PE}. We adopt priors consistent with previous eccentric analyses, including uniform priors on the initial eccentricity $e_0\in[0.001,0.5]$ and orbital anomaly $\zeta_0\in[0,2\pi]$, together with the standard priors on masses and spins.

To assess the accuracy of the surrogate in the inference setting, we compare approximately $2\times10^4$ likelihood evaluations against the reference \texttt{SEOBNRv5EHM} waveform model. The log-likelihood difference,
\begin{align}
\Delta\ln\mathcal{L}
=
\ln\mathcal{L}_{\rm SEOBNRv5EHM}
-
\ln\mathcal{L}_{\rm GWAgent},
\end{align}
is shown in Fig.~\ref{fig:supp_gw200129_corner}. The distribution is sharply peaked near zero, with median
\[
\Delta\ln\mathcal{L}\approx-1.17,
\]
corresponding to an effective signal-to-noise-ratio difference of approximately $0.4$.

The resulting posterior distributions are shown in Fig.~\ref{fig:supp_gw200129_corner}. The analysis yields
\begin{align}
e_{20\,{\rm Hz}}
=
0.099^{+0.063}_{-0.044},
\end{align}
together with source parameters consistent with previous eccentric analyses~\cite{Gupte:2024jfe,Planas:2025jny,Tang:2026jvl}. The reported eccentricity is obtained by evolving the sampled initial eccentricity through the same EOB dynamics used for waveform generation,
\begin{align}
e_{20\,{\rm Hz}}
=
e\!\left(t(f=20\,{\rm Hz})\right).
\end{align}

\section{Failure modes and human oversight}
\label{sec:app_failure_modes}

Although \GWAgent{} relies on quantitative validation against a reference simulator, human oversight remained essential throughout development. We observed several failure modes in which candidate implementations appeared plausible or successfully executed, yet failed to satisfy the intended scientific objectives. These examples illustrate both the strengths and limitations of simulation-grounded agentic workflows.

One class of failures involved implementation shortcuts that bypassed the intended validation procedure. During runtime optimization, for example, one agent cached intermediate waveform data associated with a particular validation set. Although the resulting implementation passed the timing benchmarks for that dataset, the apparent speedup disappeared when evaluated on previously unseen configurations. This behavior motivated stricter runtime tests in which cached products were removed or regenerated for every validation run.

A second class of failures involved incorrect handling of training and validation datasets. Despite explicit instructions to maintain disjoint data splits, some agent runs reported training-set performance as validation accuracy. Such errors are particularly problematic for surrogate modeling because pointwise residual metrics may appear accurate even when waveform-level generalization fails. Consequently, construction of the training and validation datasets, together with verification of the samples used for each reported metric, remained human-supervised components of the workflow.

A third failure mode involved incorrect implementation of the waveform reconstruction equations. In several runs, the generated code did not correctly implement
\[
\omega_{\rm pred}(t)
=
\omega_{\rm circ}(t)
\left[
1+\xi_\omega(t)
\right].
\]
Instead, some implementations directly integrated $\xi_\omega(t)$ as a phase increment rather than integrating the reconstructed instantaneous frequency. Although these programs executed successfully and generated output files, they accumulated phase errors of approximately $10^2$--$10^4$ rad while still terminating with ``WORKFLOW COMPLETE'' summaries. These failures motivated the Ralph-loop-like execution framework adopted in the final implementation, in which every candidate modification is executed, quantitatively evaluated, and retained only if it satisfies predefined scientific validation criteria.

Additional failure modes included overfitting in flexible ML models, excessive basis complexity with negligible validation improvement, inconsistent reporting of evaluation metrics across repeated runs, memory-exhaustion failures during basis construction, and optimization of computational speed at the expense of waveform accuracy. We also observed cases in which the agent optimized against workflow-completion objectives rather than the intended scientific metrics.

These examples demonstrate that quantitative validation substantially reduces hallucinations but does not eliminate the need for human oversight. Domain expertise remains essential for defining scientific objectives, designing validation metrics, identifying invalid shortcuts, auditing train--validation procedures, and verifying that the resulting models satisfy physically meaningful criteria rather than merely completing the prescribed workflow.

\end{document}